\documentclass[usenatbib,usegraphicx]{mn2e}

\title[HI in early-type galaxies with recent star formation]{The HI content of elliptical and lenticular galaxies with recent star formation}
\author[J. Helmboldt]{J. F. Helmboldt$^{1,2}$\thanks{E-mail: joe.helmboldt@nrl.navy.mil} \\
$^{1}$Naval Research Laboratory, Code 7213, 4555 Overlook Avenue SW, Washington, DC 20375-5351 \\
$^{2}$Department of Physics and Astronomy, University of New Mexico, 800 Yale Blvd NE, Albuquerque, NM 87131, USA}
\begin{document}

\date{Not yet submitted.}

\pagerange{\pageref{firstpage}--\pageref{lastpage}} \pubyear{2006}

\maketitle

\label{firstpage}

\begin{abstract}
As a first step toward constraining the efficiency of the star formation episodes that lead to elliptical (E) and lenticular (S0) K+A galaxies, a survey for HI within a sample of E and S0 K+A galaxies and their likely progenitors (i.e., actively star forming E and S0 galaxies) has been conducted with the NRAO Green Bank Telescope (GBT).  The sample was taken from a larger parent sample drawn from the Sloan Digital Sky Survey (SDSS).  Here, the GBT data and initial results are discussed.  Over half (19 out of 30) of all observed galaxies have detectable 21-cm emission.  It was found that both the K+A and star forming early-type (SFE) galaxies were on average more gas poor than disk galaxies at the same luminosity while being more gas rich than more typical E and S0 galaxies with detected 21-cm emission.  The gas richness of K+A galaxies appears to be similar to that of SFE galaxies.  The star formation rates and estimated star formation time scales of the SFE galaxies imply that they are capable of only marginally changing their atomic hydrogen content.  Follow-up observations are required to explore these same issues in terms of molecular gas, which is more likely to actively participate in the star formation process.  Kinematic data for the HI gas, the warm ionised gas, and the stars within the galaxies combined with the SDSS $g$ and $i$ band surface brightness profiles imply that the atomic hydrogen is most likely spatially coincident with the star forming regions within $\sim$1 kpc of the galaxies' centres.
\end{abstract}

\begin{keywords}
galaxies: elliptical and lenticular, cD -- galaxies: star-burst -- galaxies: ISM
\end{keywords}

\section{Introduction}
Identifying and understanding the processes that drive morphological changes within galaxies is essential to a general understanding of galaxy evolution.  Dissipation modulated mergers of similar sized disk galaxies have been shown to be capable of producing spheroidal galaxies similar to observed elliptical galaxies \citep[e.g.,][]{mez03,ace05}.  Many have argued that as disk galaxies move into and through galaxy clusters, processes such as gas stripping can truncate the star formation activity within the galaxies' disk components, leading to more bulge-dominated, earlier-type galaxies \citep[e.g.,][]{lar80,gun72,nul82,aba99}.  In fact, there is observational evidence that, for instance, the fraction of lenticular galaxies within clusters increases toward the clusters' centres while the fraction of disk galaxies decreases  \citep{dre97,van98,fas00}.  The stripping of gas from some disk galaxies falling into clusters has been directly observed \citep{vog04}.  For field galaxies, however, processes or events which are capable of stripping gas in a similar manner are rare.  If field galaxies evolve along the Hubble sequence to become more concentrated spheroidal galaxies, they most likely do so via some other mechanism.\par
It is within this context that so called ``K+A'' (or ``E+A'') galaxies may be extremely useful.  K+A galaxies are galaxies whose spectra have two dominant components; one that resembles that of a typical early-type galaxy or K giant star and one that resembles that of a main sequence A star.  K+A galaxies by definition have extremely weak or no nebular emission lines, implying that they are not currently forming stars.  However, the presence of an intermediate age stellar population, usually inferred from strong Balmer absorption lines, implies that these galaxies have formed stars within the last $\sim$1 Gyr.  The initial discovery of and follow-up searches for K+A galaxies identified them as belonging to a cluster population \citep{dre83,cou87}, especially at intermediate ($z\sim0.3-1$)redshifts \citep{tra04}.  However, relatively large samples of K+A galaxies culled from modern spectroscopic surveys such as the Sloan Digital Sky Survey \citep[SDSS;][]{yor00} have revealed that locally ($z^{<}_{\sim}0.2$), the fraction of K+A galaxies tends to be higher in {\it lower} density environments \citep{got05,qui04,hog06,hel07}.  Imaging data has also confirmed that these galaxies tend to be earlier-type galaxies with Sersic indexes $\sim$2-3 \citep{qui04}.  Follow-up imaging has also demonstrated that for the typical K+A galaxy, the most recent episode of star formation occurred within the centre of the galaxy \citep{yam05,helw07}.  These episodes of star formation are then capable of changing their host galaxies into more centrally concentrated galaxies not by reducing the prominence of disk components, as in the case of gas stripping, but by increasing the stellar masses/luminosities of their centres.  Taking all of this information into account, it is clear that understanding the processes that trigger and halt the star formation that leads to K+A galaxies is an important step toward understanding how field galaxies may evolve along the Hubble sequence.\par
One integral part of understanding these processes is estimating the efficiency of the star formation episodes that lead to K+A galaxies.  Among galaxies going through bursts of nuclear star formation, the most efficient star-bursts are capable of exhausting their supplies of gas in $\sim$100 Myr \citep{ken98}.  The majority of these star-bursts are also associated with mergers or galaxy-galaxy interactions \citep[e.g.,][]{lee94,san96}.  Less efficient star-bursts are associated with mergers significantly less frequently; the star formation that is found within galaxy disks, which is typically driven by internal processes, is even less efficient than these bursts \citep{ken98}.  This implies that the enhanced star formation brought about by galaxy mergers is the most efficient mode of star formation found within galaxies.  Estimating the efficiency of the star formation that leads to K+A galaxies is then crucial to constraining the processes that may be driving that star formation.\par
\begin{figure*}
\includegraphics[scale=0.75]{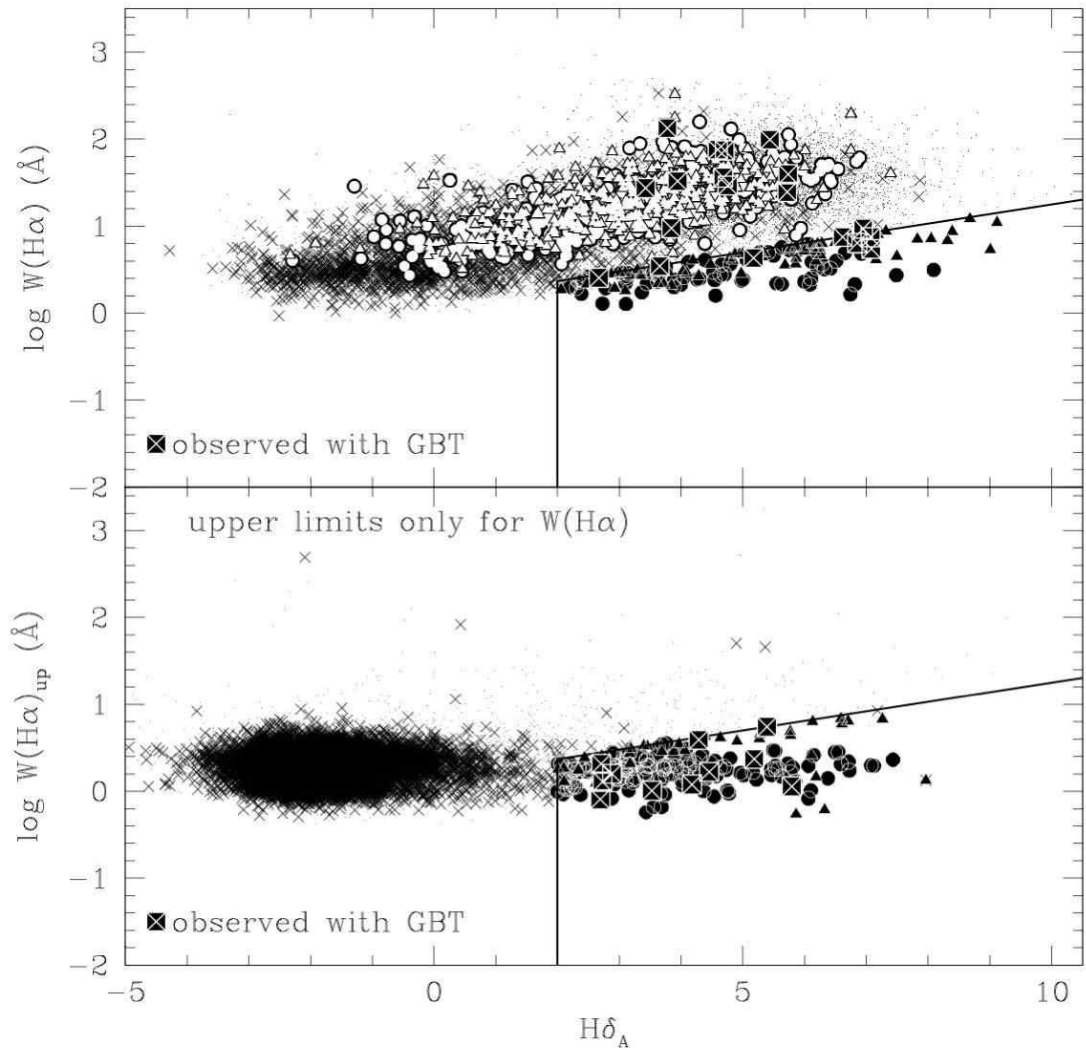}
\caption{From the fourth data release (DR4) of the SDSS and the measurements of \citet{tre04}, the H$\alpha$ emission line equivalent width, $W(H\alpha)$ ({\it emission} is positive), versus the strength of the H$\delta$ absorption line, $H\delta_{A}$ ({\it absorption} is positive), for galaxies with $\geq 3 \sigma$ detections of H$\alpha$ and $m_{r}<$16 (upper).  For galaxies with no significant detection of H$\alpha$, the upper limit for $W(H\alpha)$ is plotted in the lower panel.  Early-type galaxies are represented by $\times$'s; star forming early-type (SFE) galaxies are represented by white circles for elliptical and lenticular galaxies and white triangles for early-type spiral galaxies.  Similarly, elliptical and lenticular K+A galaxies are represented by black circles and spiral K+A galaxies are represented by black triangles.  Galaxies that were observed with the GBT are highlighted as boxes.  In both panels, the definition of K+A galaxies used by \citet{hel07} is illustrated by the solid lines.}
\label{classdef}
\end{figure*}

To obtain a statistical estimate of the star formation efficiency, one needs to measure the amount of cold gas contained within a sample of K+A galaxies and within a sample of their actively star forming progenitors.  A sample of 335 star forming elliptical (E) and lenticular (S0) galaxies with $m_{r}<16$ taken from the fourth data release of the SDSS has been identified by \citet{hel07} as most likely being a sample of the progenitors of morphologically similar (i.e. E and S0) K+A galaxies.  These star forming early-type galaxies, or SFE galaxies, were identified as actively forming stars by their emission line ratios using the emission line fluxes measured by \citet{tre04}.  They were also morphologically classified by visual inspection of their SDSS $g$-band images down to a limiting $r$-band magnitude of 16.  A sample of 253 E and S0 K+A galaxies with $m_{r}<16$ were also selected from the SDSS.  To maximise the K+A sample size, a less stringent definition was used than has been used by some authors \citep[e.g.,][]{zab96,tra04,got03}, but which is similar to that used by \citet{qui04}.  Formally, it was required that a K+A galaxy have $H\delta_{A}>2\mbox{ \AA}$ and log $W(H\alpha)<0.11H\delta_{A}+0.15$, where $W(H\alpha)$ is the H$\alpha$ emission line equivalent width in units of $\mbox{\AA}$ measured by \citet{tre04} and $H\delta_{A}$ is the spectral index defined by \citet{wor97} to measure the strength of the H$\delta$ absorption line, also in units of $\mbox{\AA}$.  The values for $H\delta_{A}$ were taken from \citet{tre04} and include their corrections for H$\delta$ emission.  This definition was empirically derived using the location of all actively star forming galaxies and all quiescent early-type galaxies within the SDSS with $m_{r}<16$ (see Fig.\ \ref{classdef}).\par

It was found that the distributions of masses as traced by stellar velocity dispersion were nearly identical for star forming E and S0 galaxies and E and S0 K+A galaxies.  The fractions of these two types of galaxies among all SDSS galaxies with $m_{r}<$16 also depend on environment in nearly the same way.  Modelling of the star formation histories of the star forming E and S0 galaxies implies that their properties are consistent with episodes of star formation that last about 200 Myr on average.  This time scale is short enough for them to become K+A galaxies.  The model prediction for the distribution of H$\delta$ absorption line strengths for the star forming E and S0 galaxies as they become K+A galaxies is nearly identical to that observed for the actual E and S0 K+A galaxies.  Therefore, in addition to being morphologically similar, the star forming E and S0 galaxies and the E and S0 K+A galaxies appear to be linked in a clear evolutionary sequence.\par
The star forming E and S0 galaxies and the E and S0 K+A galaxies from the \citet{hel07} sample provide the opportunity to explore the efficiency of the star formation episodes that likely lead to elliptical and lenticular K+A galaxies.  This is unique to the \citet{hel07} sample because (i) there is evidence that these two particular samples, which are relatively large, are evolutionarily linked (ii) there are few other known actively star forming elliptical galaxies at low redshift \citep{fuk04} and (iii) previous H{\sc i} measurements for a few actively star forming S0 galaxies are confined to only the most gas-rich objects \citep{pog93}.  As a first step toward constraining the efficiency of the star formation episodes that lead to E and S0 K+A galaxies, a survey for neutral hydrogen has been conducted with the NRAO Green Bank Telescope (GBT) within a subset of star forming and K+A elliptical and lenticular galaxies.  These observations will provide a first look at the cold gas content of these objects and will allow for a comparison of the gas richness of each of the two galaxy classes to each other and to other types of galaxies.  This information will be used to select candidates for follow-up observations aimed at detecting molecular gas which will provide a much better estimate of the amount of "fuel" for star formation that is available within both classes of galaxies.  In this paper, the observations, data, and general HI properties are presented (\S 2 and \S 3) and future follow-up observations are discussed (\S 4).
 
\begin{table*}
\centering
\caption{Observations}
\begin{tabular}{ccccccc}
\hline
Name & UT Date & Exp. Time (s) & Morph. Type & Spec. Type & $V_{r}$ (km s$^{-1}$) & Alt. Name \\
\hline
J003823.71$+$150222.56 &    2006-08-19 & 3438 & S0 & SFE & 5384 &  UGC 00386 \\
J013214.68$-$090635.24 &    2006-08-19 & 3438 &  E & SFE & 5311 & $\cdots$ \\
J013730.83$-$085307.73 &    2006-08-19 & 1146 & S0 & K+A & 1797 & MCG -02-05-026 \\
J015432.72$-$004612.40 &    2006-08-16 & 4585 &  E & K+A & 4819 & $\cdots$ \\
J024032.84$-$080851.65 &    2006-08-16 & 1376 & S0 & K+A & 1340 & NGC 1047 \\
J031117.74$-$080448.04 &    2006-08-16 & 2293 &  E & SFE & 4004 & $\cdots$ \\
J031651.20$+$411520.87 & 2006-08-19/20 & 4586 &  S0 & K+A & 1665 & $\cdots$ \\
J032324.54$+$402118.36 & 2006-08-19/20 & 5733 &  S0 & K+A & 2216 & $\cdots$ \\
J080142.49$+$251425.08 &    2006-08-20 & 2294 & S0 & K+A & 4685 & CGCG 118-049 \\
J083228.06$+$523622.32 &    2006-08-19 & 2294 &  E & SFE & 5094 & MRK 0091 \\
J090244.63$+$311626.04 &    2006-08-20 & 2294 &  E & SFE & 4145 & CGCG 150-062 \\
J102757.13$+$603802.77 &    2006-08-19 & 2294 &  E & K+A & 1298 & MCG +10-15-083 \\
J103801.68$+$641559.00 &    2006-08-19 & 2294 &  E & SFE & 1700 & UGC 05776 \\
J111059.99$+$525917.88 &    2006-08-19 & 2294 &  E & K+A &  810 & $\cdots$ \\
J113744.40$+$540244.52 &    2006-08-19 & 2293 &  E & K+A &  907 & $\cdots$ \\
J115143.20$+$595009.59 & 2006-08-19/20 & 4587 &  E & K+A & 3495 & SBS 1149+601\\
J121024.49$+$131014.16 &    2006-08-19 & 2294 &  E & K+A & 1691 & KUG 1207+134 \\
J121458.09$+$525639.84 & 2006-08-19/20 & 4587 &  E & SFE & 5441 & CGCG 269-046 \\
J130658.07$+$521526.64 &    2006-08-19 & 2294 &  E & K+A & 4753 & MCG +09-22-012 \\
J133253.05$-$011531.14 &    2006-08-19 &  764 &  E & K+A & 3592 & CGCG 017-019 \\
J140058.32$+$553405.16 &    2006-08-19 & 2294 & S0 & K+A & 1852 & $\cdots$ \\
J140123.99$+$364800.35 &    2006-08-19 & 2294 & S0 & SFE & 2706 & MRK 0465 \\
J140820.65$+$505240.44 &    2006-08-19 & 2293 &  E & K+A & 2401 & $\cdots$ \\
J142054.96$+$400715.59 &    2006-08-19 & 4588 &  E & SFE & 5273 & CGCG 219-071 \\
J144425.44$+$415140.69 & 2006-08-19/20 & 5735 &  E & SFE & 5300 & $\cdots$ \\
J150747.75$+$011731.38 &    2006-08-19 & 2293 &  E & K+A & 2099 & CGCG 021-011 \\
J160723.27$+$414232.04 & 2006-08-19/20 & 8029 &  E & SFE & 5453 & CGCG 223-041 \\
J210729.75$+$092113.82 &    2006-08-16 & 2292 & S0 & K+A & 4136 & $\cdots$ \\
J222730.71$-$093953.97 &    2006-08-14 &  153 & S0 & SFE & 1700 & $\cdots$ \\
J225304.56$+$010839.95 &    2006-08-16 & 2292 &  E & K+A & 4655 & NGC 7402 \\
\hline
 \end{tabular}
 \label{obstab}
 \end{table*}
 
\begin{figure*}
\includegraphics[scale=0.92]{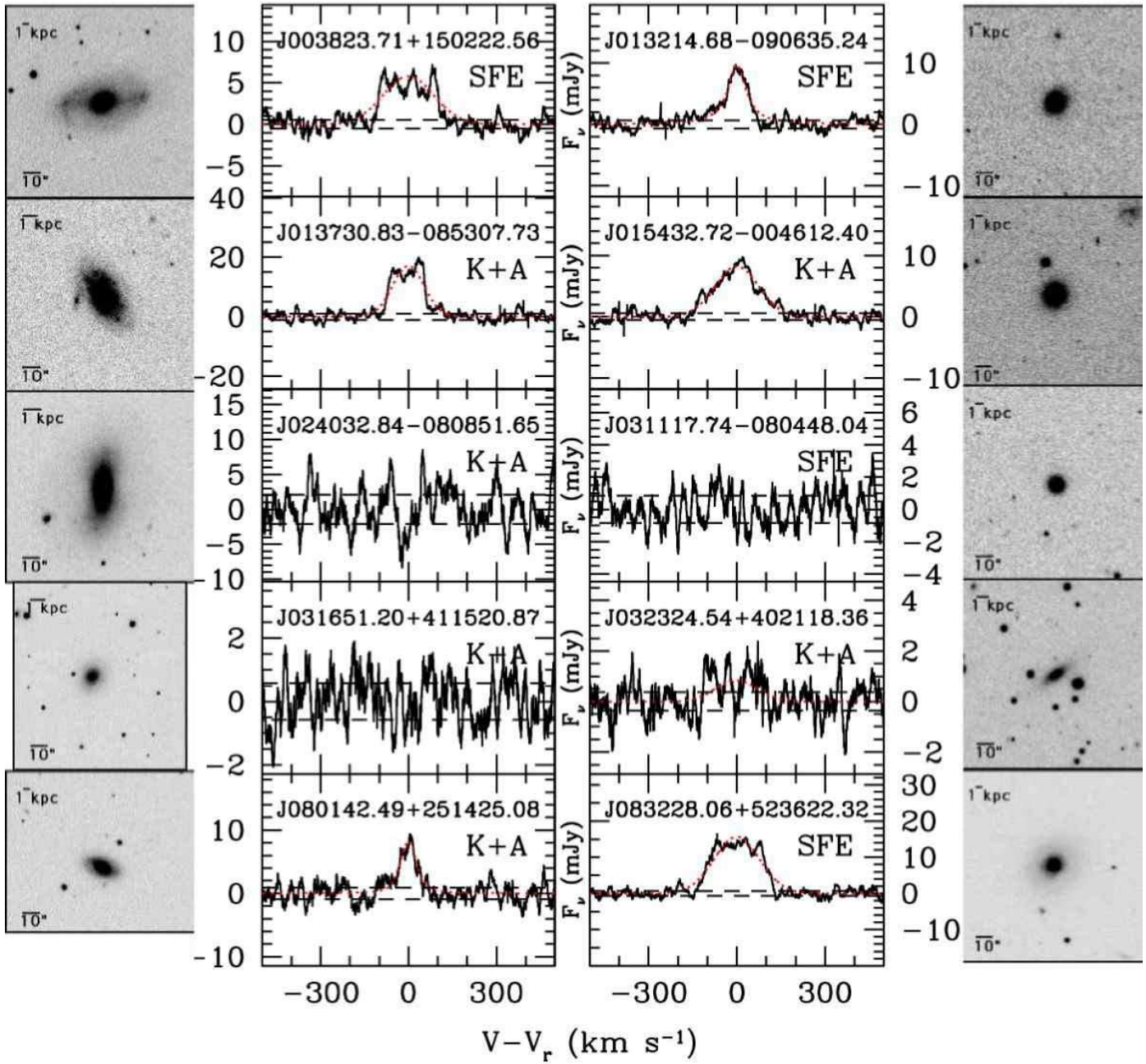}
\caption{The continuum subtracted GBT spectra for the first 10 galaxies in Table \ref{obstab} in units of mJy.  Each spectrum is accompanied by its galaxy's SDSS $g$-band image.  For each galaxy spectrum, the measured rms value for the continuum is marked above and below $F_{\nu}=0$ with dashed lines.  For galaxies with detected 21-cm flux density (see \S 2), a Voigt profile fit to the emission line is plotted as a red dotted line.  The line centres and velocity widths derived from these fits are given in Table \ref{hitab}.}
\label{specs1}
\end{figure*}

\begin{figure*}
\includegraphics[scale=0.92]{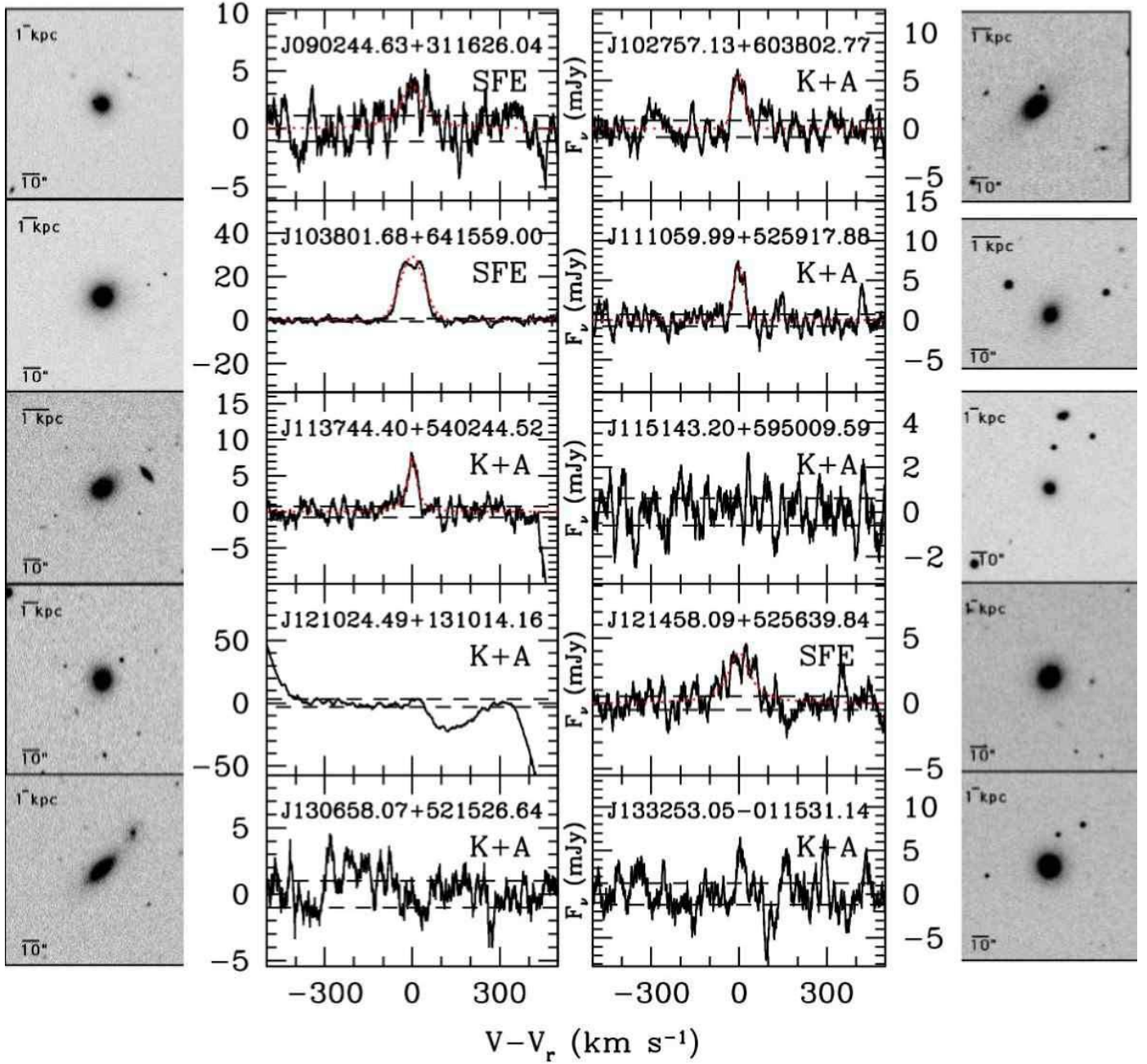}
\caption{The same as Fig.\ \ref{specs1}, but for the next 10 galaxies in Table \ref{obstab}.}
\label{specs2}
\end{figure*}

\begin{figure*}
\includegraphics[scale=0.92]{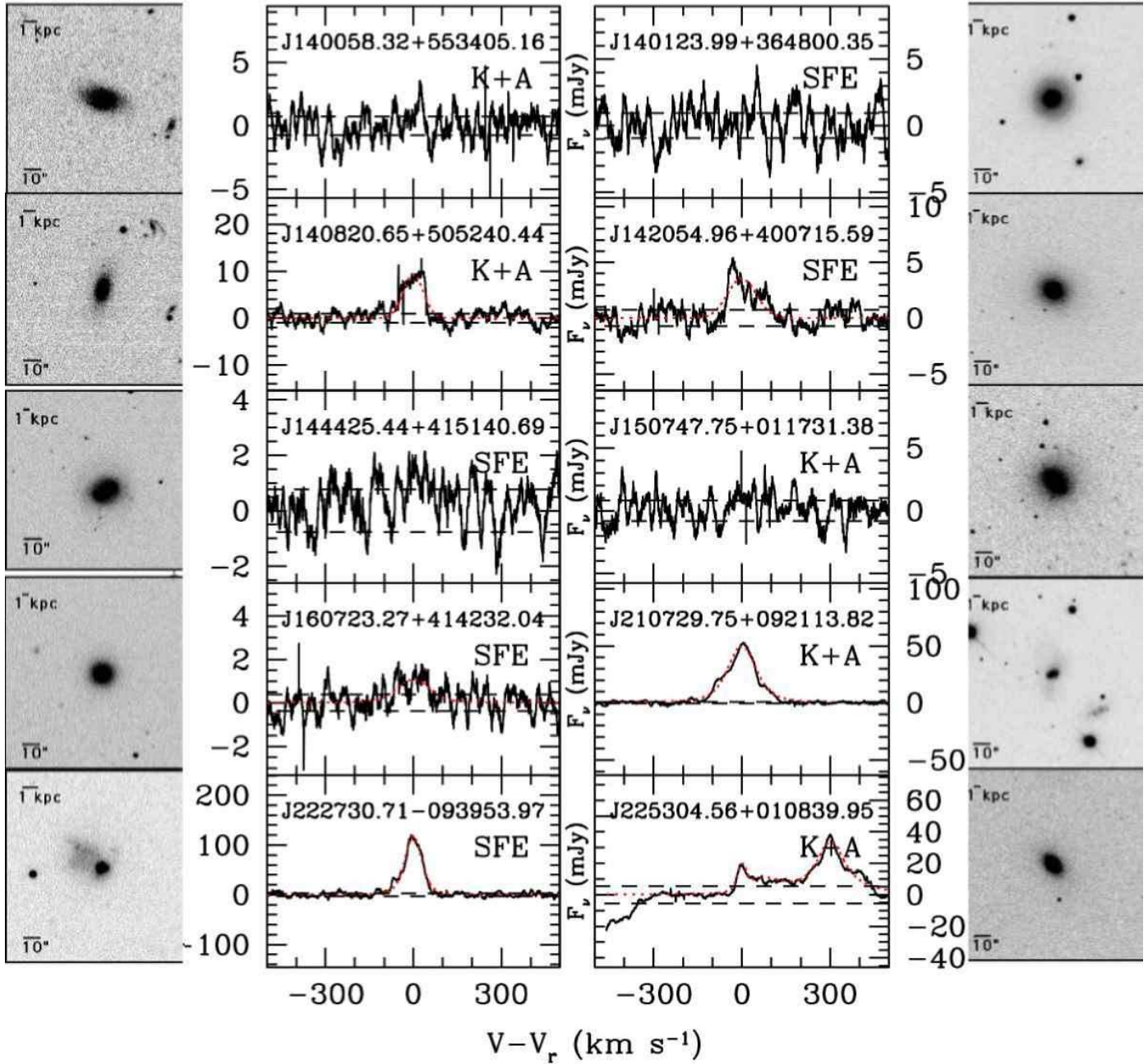}
\caption{The same as Fig.\ \ref{specs1}, but for the last 10 galaxies in Table \ref{obstab}.}
\label{specs3}
\end{figure*}

\begin{table*}
 \centering
 \caption{Derived properties}
 \begin{tabular}{ccccccccc}
 \hline
 Name & rms (mJy) & $M_{HI}$ (M$_{\odot}$) & $W_{50}$ (km/s) & $V_{r}$(HI) (km/s) &
 $M_{B}$ & $M_{R}$ & $W_{50}$(stars) & $W_{50}$(H$\alpha$) \\
 \hline
J003823.71$+$150222.56 & 0.51 & $1.46\pm0.05\times 10^{9}$ & 201.1 & 5358 & -19.58 & -20.98 & 220 & 73.5 \\
J013214.68$-$090635.24 & 0.66 & $8.3\pm0.4\times 10^{8}$ & 73.8 & 5258 & -17.96 & -18.92 & 138 & 47.7 \\
J013730.83$-$085307.73 & 1.05 & $3.3\pm0.1\times 10^{8}$ & 127.0 & 1785 & -16.12 & -17.14 & 48 & $\cdots$ \\
J015432.72$-$004612.40 & 0.54 & $1.66\pm0.04\times 10^{9}$ & 172.0 & 4734 & -18.19 & -19.48 & 40 & $\cdots$ \\
J024032.84$-$080851.65 & 2.09 & $<2.68\times 10^{7}$ & $\cdots$ & $\cdots$ & -16.76 & -18.07 & 150 & $\cdots$ \\
J031117.74$-$080448.04 & 0.83 & $<9.62\times 10^{7}$ & $\cdots$ & $\cdots$ & -17.07 & -18.08 & 68 & 44.8 \\
J031651.20$+$411520.87 & 0.58 & $<1.15\times 10^{7}$ & $\cdots$ & $\cdots$ & -15.79 & -17.21 & 137 & $\cdots$ \\
J032324.54$+$402118.36 & 0.36 & $3.9\pm0.4\times 10^{7}$ & 166.3 & 2106 & -15.84 & -17.06 & 32 & $\cdots$ \\
J080142.49$+$251425.08 & 0.89 & $6.3\pm0.4\times 10^{8}$ & 65.7 & 4660 & -18.28 & -19.65 & 79 & $\cdots$ \\
J083228.06$+$523622.32 & 0.69 & $3.47\pm0.06\times 10^{9}$ & 179.7 & 5083 & -19.31 & -20.72 & 524 & 91.1 \\
J090244.63$+$311626.04 & 1.1 & $3.2\pm0.5\times 10^{8}$ & 87.8 & 4129 & -17.99 & -19.29 & 162 & 76.7 \\
J102757.13$+$603802.77 & 0.87 & $2.7\pm0.3\times 10^{7}$ & 46.6 & 1269 & -15.01 & -16.03 & 79 & $\cdots$ \\
J103801.68$+$641559.00 & 0.67 & $3.87\pm0.05\times 10^{8}$ & 96.2 & 1698 & -17.10 & -18.34 & 60 & 83.8 \\
J111059.99$+$525917.88 & 0.74 & $9.0\pm0.8\times 10^{6}$ & 42.0 & 828.6 & -13.84 & -14.79 & 75 & 36.2 \\
J113744.40$+$540244.52 & 0.78 & $1.4\pm0.1\times 10^{7}$ & 37.3 & 902.3 & -14.07 & -15.09 & 38 & 40.9 \\
J115143.20$+$595009.59 & 0.61 & $<5.39\times 10^{7}$ & $\cdots$ & $\cdots$ & -16.43 & -17.92 & 94 & $\cdots$ \\
J121024.49$+$131014.16 & 3.12 & $<6.38\times 10^{7}$ & $\cdots$ & $\cdots$ & -15.47 & -16.42 & 108 & $\cdots$ \\
J121458.09$+$525639.84 & 0.53 & $6.0\pm0.4\times 10^{8}$ & 85.7 & 5471 & -18.62 & -20.02 & 188 & 58.0 \\
J130658.07$+$521526.64 & 0.99 & $<1.62\times 10^{8}$ & $\cdots$ & $\cdots$ & -17.59 & -18.96 & 37 & $\cdots$ \\
J133253.05$-$011531.14 & 1.22 & $<1.14\times 10^{8}$ & $\cdots$ & $\cdots$ & -18.02 & -19.12 & 19 & $\cdots$ \\
J140058.32$+$553405.16 & 0.73 & $<1.79\times 10^{7}$ & $\cdots$ & $\cdots$ & -15.26 & -16.41 & 125 & $\cdots$ \\
J140123.99$+$364800.35 & 0.93 & $<4.92\times 10^{7}$ & $\cdots$ & $\cdots$ & -18.00 & -19.20 & 112 & 64.7 \\
J140820.65$+$505240.44 & 0.99 & $2.3\pm0.1\times 10^{8}$ & 80.2 & 2148 & -15.89 & -17.13 & 87 & $\cdots$ \\
J142054.96$+$400715.59 & 0.73 & $5.4\pm0.5\times 10^{8}$ & 113.2 & 5299 & -18.91 & -20.17 & 129 & 83.8 \\
J144425.44$+$415140.69 & 0.76 & $<1.54\times 10^{8}$ & $\cdots$ & $\cdots$ & -18.43 & -19.85 & 186 & 54.2 \\
J150747.75$+$011731.38 & 0.81 & $<2.56\times 10^{7}$ & $\cdots$ & $\cdots$ & -16.61 & -17.86 & 110 & $\cdots$ \\
J160723.27$+$414232.04 & 0.38 & $1.8\pm0.3\times 10^{8}$ & 134.5 & 5391 & -18.45 & -19.71 & 135 & 53.6 \\
J210729.75$+$092113.82 & 0.75 & $5.15\pm0.04\times 10^{9}$ & 114.5 & 4170 & -17.62 & -18.17 & 29 & 57.5 \\
J222730.71$-$093953.97 & 3.02 & $1.28\pm0.02\times 10^{9}$ & 68.9 & 1675 & -15.83 & -16.28 & 636 & 44.3 \\
J225304.56$+$010839.95 & 5.57 & $1.4\pm0.3\times 10^{9}$ & 34.7 & 4575 & -17.93 & -19.22 & 54 & $\cdots$ \\
\hline
\end{tabular}
\label{hitab}
\end{table*}

\section{Sample selection, observations, and data reduction}

\begin{figure}
\includegraphics[scale=0.42]{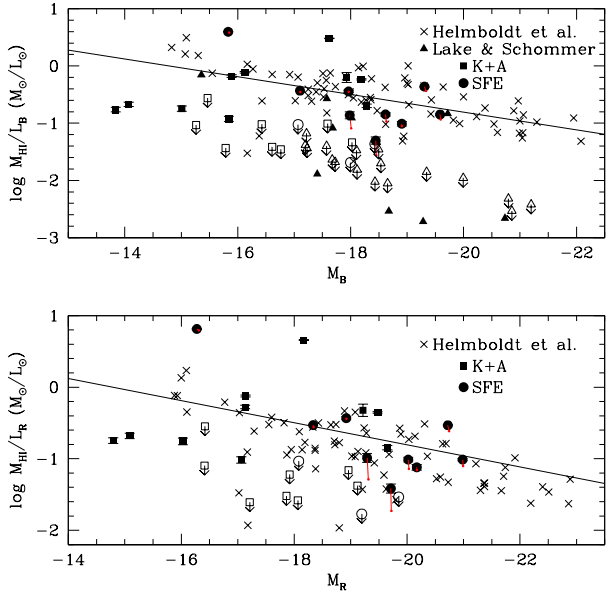}
\caption{The gas-to-light ratio versus luminosity in the B (upper) and R (lower) bands for the SFE (circles) and K+A (squares) galaxies determined using the SDSS $g$ and $r$ band Petrosian magnitudes and the conversions given in \citet{smi02}.  For both types of galaxies, upper limits are represented by open points with arrows for galaxies with no detected 21-cm emission.  In both panels, the red lines indicate the path each SFE galaxies will traverse after 200 Myr of star formation assuming that only HI is consumed by the star formation and stellar mass-to-light ratios of 2.27 and 1.56 in the B and R bands respectively (see \S 3.1).  Also plotted is the data from \citet{hel04} (represented by $\times$'s); the solid lines are linear fits to these data.  The triangles represent the data for E and S0 galaxies taken from \citet{lak84}.}
\label{gaslight}
\end{figure}

\begin{figure}
\includegraphics[scale=0.42]{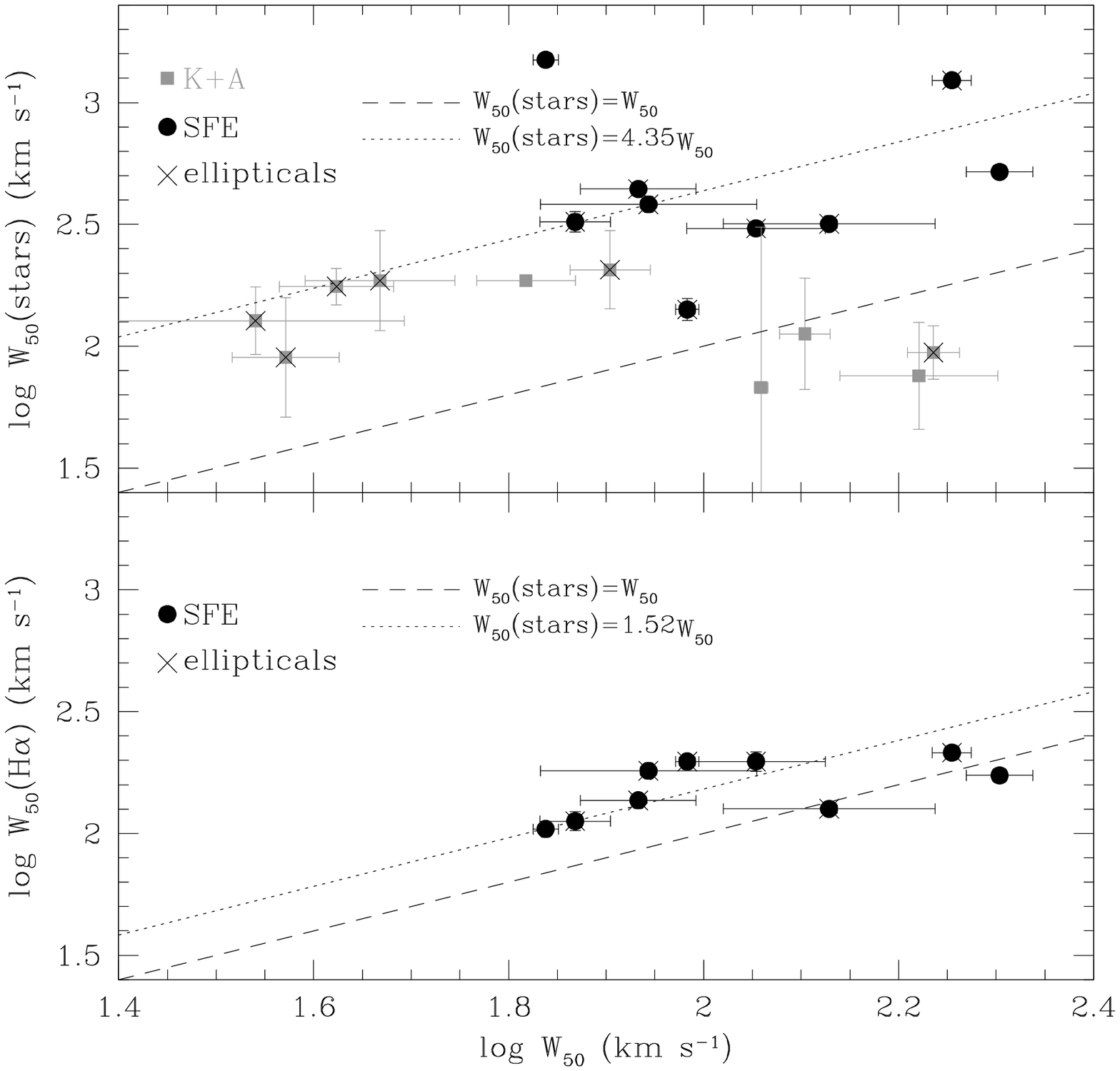}
\caption{The full width at half power of the stellar velocity distribution, $W_{50}$(stars) (upper), and the H$\alpha$ emission line, $W_{50}$H($\alpha$) (lower), measured from the SDSS spectra versus the 21-cm line velocity width, $W_{50}$, measured from the GBT spectra.  Only the SFE galaxies are included in the lower panel since any H$\alpha$ emission detected within the K+A galaxies in not likely linked to current star formation.  In both panels, black points represent SFE galaxies and grey boxes represent K+A galaxies.  Points representing elliptical galaxies are flagged with $\times$'s.  The dashed lines represent the case where the SDSS-measured velocity widths match the values of $W_{50}$; the dotted lines represent the median ratio of stellar/ionised gas velocity width to the HI velocity width for SFE galaxies only, which is equal to 4.35 for the stellar velocity and 1.52 for the ionised gas velocity.}
\label{vwidths}
\end{figure}

As discussed above, all galaxy targets were selected from the \citet{hel07} sample of elliptical and lenticular star forming and K+A galaxies.  The sample was chosen to be large enough as to be representative of the parent sample but was also chosen to be nearby enough that relatively low HI mass detection limits could be reached within a reasonable ($<$3 hours per source) amount of observing time.  With the excellent sensitivity of the GBT, a $3\sigma$ detection limit of 10$^{8}$ M$_{\odot}$ could be reached for 41 of the \citet{hel07} galaxies within the chosen time constraints.  This is a full order of magnitude lower than detection limits quoted in previous searches for HI in K+A galaxies using other instruments, owing both to the sensitivity of the GBT and to the less stringent K+A definition used which allowed for the identification of K+A galaxies that are more nearby than those that have previously been searched for HI \citep[e.g.,][]{cha01,buy06}.  The NASA/IPAC Extragalactic Database (NED) was search near each of these potential targets to eliminate any sources that had one or more galaxies nearby that would be contained within the GBT beam and were at radial velocities (a difference of $<$300 km s$^{-1}$) that would make it difficult to determine the source of any detected 21-cm emission.  This eliminated four potential targets.\par
Data was obtained with the GBT for 30 of the remaining galaxies in August, 2006 (see Table \ref{obstab} for a summary), with 7 targets being excluded due to time constraints and their proximity to the sun during the observing run.  For all galaxies, the GBT spectrometer was used with a total bandwidth of 12.5 MHz, 16,384 channels, and a central frequency of 1420.405(1+z)$^{-1}$ MHz where z is the redshift measured from the SDSS optical spectrum of each target.  The exposure time used was adjusted for each target by monitoring the rms noise of its spectrum in real time as the galaxy was observed in position switching mode in intervals of 10 minutes on and off the source with the goal of reaching a $3\sigma$ detection limit of $10^{8}$ M$_{\odot}$ assuming H$_{\circ}=$70 km s$^{-1}$ Mpc$^{-1}$ for a velocity width of 200 km s$^{-1}$, the median H$\alpha$ velocity width for the SFE galaxies.  In position switching mode, an exposure is taken while pointing at the object immediately followed by another exposure, usually of equal duration, of a blank part of the sky.  For these observations, the "blank" sky exposure was obtained by slewing to a position 1$^{\circ}$ away from the target in both right ascension and declination.  For exposures on and off the source that are of equal length, the antenna temperature is given by
\begin{equation}
T_{\nu,A} = T_{sys} \frac{S_{\nu,on} - S_{\nu,off}}{S_{\nu,off}}
\end{equation}
where observations of bright radio sources of known flux density at 1400 MHz were used to convert the system temperature, $T_{sys}$, into units of flux density.  These computations were done at the telescope with the software package GBTIDL.  After combining all on and off target exposures for each source, the final calibration of the spectra was done within IRAF using customised scripts.  After calibrating them, the spectra were smoothed with a 100 channel wide boxcar to yield an effective channel width of approximately 16.7 km s$^{-1}$.  The final step in the data reduction involved fitting a cubic spline function to the continuum of each spectrum within IRAF while interactively adjusting the location and size of the fitting window(s) and the number of spline segments used (typically between 10 and 15).  The final calibrated, continuum subtracted spectra are displayed in Fig.\ \ref{specs1}-\ref{specs3} along with the SDSS $g$-band images.\par
For each galaxy, the peak flux density was measured within $\pm$500 km s$^{-1}$ of the SDSS-measured radial velocity from the continuum subtracted spectrum and was compared to the rms flux density measured outside this 1,000 km s$^{-1}$ window.  For those sources where the peak flux density was more than five times the rms, the integrated flux of the emission line, $S_{int}$, was computed and the velocity width was roughly estimated to be $W_{50} \approx S_{int}/S_{peak}$.  The error in the integrated flux was then computed using the rms measured outside the 21-cm emission line window and assuming the emission line spans 2$W_{50}/\Delta v$ channels where $\Delta v$ is the width of a single channel in units of km s$^{-1}$.  For these galaxies, the median value of the estimate of $W_{50}$ is approximately 75 km s$^{-1}$.  Using this fact, a 3$\sigma$ upper limit for the integrated flux was computed for the remaining sources assuming $W_{50}=$ 75 km s$^{-1}$. These upper limits were compared to the integrated fluxes measured from the continuum subtracted spectra within $\pm$500 km s$^{-1}$ of the expected line centre.  Those galaxies whose integrated fluxes were larger than this upper limit were considered detections and the errors in the integrated fluxes were computed as above.  For each galaxy with an HI detection, the HI mass was computed using the integrated flux and assuming H$_{\circ}=$70 km s$^{-1}$ Mpc$^{-1}$.\par
To measure the location of the centre of the HI line, as well as to obtain a better measurement of the full width at half power, $W_{50}$, a Voigt profile was fit to each HI emission line.  Rather than being motivated by physical reasons, the choice of the Voigt profile was made to provide a more flexible function than a simpler profile (e.g., a Gaussian) because of the somewhat irregular shapes of some of the emission lines.  The fitting of Voigt profiles also allows for reliable measurements of the centres and velocity widths of the emission lines for those galaxies with relatively weak detections (e.g., J090244.63$+$311626.04).  For all but one galaxy, J140820.65$+$505240.44, the radial velocity of the HI lines estimated in this manner agreed with the radial velocities measured from the SDSS spectra within $\pm W_{50}$/2.  For J140820.65$+$505240.44, the HI radial velocity is about 250 km s$^{-1}$ smaller than the SDSS-measured radial velocity and there are no obvious optical companions within the area of the GBT beam.  This velocity discrepancy may be the result of a significant amount of HI gas that is currently being deposited within this galaxy.  Follow-up radio frequency spectral imaging is required to adequately address this issue.    A second galaxy, J225304.56$+$010839.95, has a companion galaxy that is nearby both in position on the sky and in radial velocity.  This companion, NGC 7401, is at a radial velocity of about 370 km s$^{-1}$ larger than that of J225304.56$+$010839.95 according to NED and was also detected in 21-cm emission.  As can be seen from the spectrum plotted in Fig.\ \ref{specs3}, the emission lines from these two galaxies are somewhat blended.  Separate Voigt profiles were fit simultaneously to effectively de-blend the two line profiles so that the line centres and velocity widths could be estimated for J225304.56$+$010839.95 and NGC 7401 separately.  For the measurement of the HI mass of J225304.56$+$010839.95, the emission line window was adjusted by eye to isolate its emission from that of NGC 7401.\par
All derived parameters discussed above are listed in Table \ref{hitab}.  Galaxies whose integrated fluxes were less than their estimated 3$\sigma$ upper limits were considered non-detections; only the HI mass upper limits and rms values are listed in Table \ref{hitab} for these galaxies.  Overall, 19 of the 30 targets had detected 21-cm emission; nearly all (9 out of 12) SFE galaxies had detectable HI; a little more than half (10 out of 18) K+A galaxies had detected emission from HI.  Among all 30 galaxies, roughly equal fractions of elliptical (13 out of 20) and lenticular (5 out of 8) galaxies had detected 21-cm emission.\par
For one galaxy, J121024.49$+$131014.16, the GBT observations were taken when the object was relatively close to the sun.  As a result, the true shape of the continuum was not recovered using the on/off technique given by equation (1), and the resulting irregular continuum could not be adequately subtracted as evidenced by the spectrum plotted in Fig.\ \ref{specs2}.  The upper limit for the HI mass of this galaxy should therefore be taken only as a rough estimate.

\section{Results and discussion}
\subsection{Gas richness}
With HI detections for nearly two thirds of the observed SFE and K+A galaxies and relatively stringent upper limits on the HI mass for the remaining galaxies, a comparison of the gas richness of these galaxies to that of other galaxies can be made.  To this end, the samples of \citet{hel04} and \citet{lak84} were chosen as comparison samples.  The \citet{hel04} sample consists of 69 galaxies drawn from the HI Parkes All Sky Survey \citep[HIPASS;][]{bar01} that were imaged using B, R, and narrow-band H$\alpha$ filters and are predominantly spiral and irregular galaxies.  The \citet{lak84} sample consists of 28 faint ($M_{B}>-20$ for H$_{\circ}=50$ km s$^{-1}$ Mpc$^{-1}$) E and S0 galaxies observed with the 305-m telescope of the Arecibo Observatory, 12 of which were detected in 21-cm emission.  For the \citet{hel04} sample, the so called gas-to-light ratio, or $M(HI)/L$ was computed in the B and R bands using the integrated 21-cm flux from the HIPASS spectra.  Since only B-band optical magnitudes were available for the majority of the \citet{lak84} galaxies, only $M(HI)/L_{B}$ was computed for these galaxies using the published values for $L_{B}$ and $M(HI)$.  For the SFE and K+A galaxies, the SDSS $g$ and $r$ band Petrosian magnitudes were used with the conversions given by \citet{smi02} along with the GBT data to compute values or upper limits for $M(HI)/L_{B}$ and $M(HI)/L_{R}$.  The gas-to-light ratio is plotted as a function of luminosity in both bands in the panels of Fig.\ \ref{gaslight} with linear fits to the \citet{hel04} data.  From these plots, it is evident that both SFE and K+A galaxies are on average more gas poor than typical disk galaxies at the same luminosity.  About 75\% of both SFE and K+A galaxies lie below the lines fit to the \citet{hel04} data with the upper limits for all galaxies with no detected 21-cm emission lying below these lines.  In contrast, for those galaxies with detected 21-cm emission, the SFE and K+A galaxies appear to be more gas rich than the E and S0 galaxies of \citet{lak84}.  These results imply that the HI content of the SFE and K+A galaxies is on average somewhere in between what is typical for disk galaxies and the average value for E and S0 galaxies and are consistent with what was found by \citet{buy06}.  The same was found to be true for the distributions of stellar mass and velocity dispersion for the parent SFE and K+A samples by \citet{hel07}.\par
The results summarised above imply that the gas richness of the SFE galaxies is on average relatively similar to that of the K+A galaxies.  Does this then imply that the star formation that leads to K+A galaxies is very inefficient?  This is not necessarily true.  It is more likely that the relative amount of molecular and not atomic hydrogen within these galaxies will provide a true estimate of the average star formation efficiency since it is more likely that molecular gas will more actively participate in the star formation process.  However, the fact that the gas-to-light ratios for the SFE and K+A galaxies are quite similar at the same luminosity might imply that they are not evolutionarily linked as was concluded by \citet{hel07} since one would naively expect the K+A galaxies to be more gas poor on average.  Yet, even if the atomic hydrogen was the primary fuel for star formation within these galaxies, their moderate ($\sim$2 M$_{\odot}$ yr$^{-1}$) star formation rates (SFRs) and relatively short star formation times scales \citep[$\sim$200 Myr;][]{hel07} would cause their gas content to change by a relatively small amount.  This is illustrated in Fig.\ \ref{gaslight} where we have re-computed the gas-to-light ratios for the SFE galaxies using the SFR per unit B and R band luminosities, $\Psi_{B}$ and $\Psi_{R}$, computed using the H$\alpha$ emission line flux and the SDSS $g$ and $i$ band "fibre" magnitudes (i.e., the magnitudes measure within a 3 arcsec aperture).  This was done assuming that the gas-to-light ratio of each galaxy is reduced according to
\begin{equation}
\frac{M(HI)}{L} = \frac{M(HI)_{0} - L_{0} \Psi t_{co}}{L_{0}(1 + \Psi t_{co} \Upsilon_{\ast}^{-1})}
\end{equation}
where $M(HI)_{0}$ and $L_{0}$ are the initial HI mass and luminosity, $t_{co}$ is the star formation "cut-off" time which was assumed to be 200 Myr, and $\Upsilon_{\ast}$ is the stellar mass-to-light ratio.  Using the stellar masses measured by \citet{kau03}, the median values for $\Upsilon_{\ast}$ for the SFE galaxies of \citet{hel07} were determined to be 2.27 and 1.52 in the B and R bands respectively and were assumed for all SFE galaxies for this computation.  Using the re-computed values of $M(HI)/L_{B}$ and $M(HI)/L_{R}$, the path each SFE galaxy would take is plotted as a red line in Fig.\ \ref{gaslight} and show that while the gas-to-light ratios of the most gas poor SFE galaxies will change significantly, the overall gas richness of the SFE galaxy sub-sample was changed relatively little.  Therefore, it appears that the similarity between the gas-to-light ratios for SFE and K+A galaxies does not rule out the scenario in which SFE galaxies evolve into K+A galaxies.  Future follow-up observations in the millimetre regime aimed at detecting emission lines from CO to measure the relative amounts of molecular gas within SFE and K+A galaxies are required to both test the validity of this proposed scenario and to estimate the typical star formation efficiency (or, amount of molecular gas consumption) for these systems.

\subsection{The location of the HI}
The results discussed above imply that the SFE galaxies are capable of only using up a relatively small fraction of their neutral hydrogen via star formation.  Since the single-dish GBT observations provide no spatial information, one may then question whether the majority of the HI is spatially coincident with the regions of star formation, or if it typically extends substantially beyond these regions.  The kinematic information available from both the SDSS and GBT spectra can provide some insight into this issue.  Using the IDL program {\it vdispfit} written by D. Schlegel, the line-of-sight (LOS) stellar velocity dispersion was measured for each galaxy using its SDSS spectrum.  The {\it vdispfit} routine determines the best-fitting velocity dispersion and the 1$\sigma$ error in that dispersion by cross-correlating each spectrum with several template spectra that have been broadened by various Gaussian velocity distributions while masking regions of the spectrum that may contain emission lines.  Velocity widths for the H$\alpha$ emission line for all galaxies with $>5\sigma$ detections of that line were also obtained from \citet{tre04}.  In Fig.\ \ref{vwidths}, the full width at half power of the LOS stellar velocity distribution, $W_{50}$(stars), and of the H$\alpha$ emission line, $W_{50}$(H$\alpha$), are plotted as functions of $W_{50}$, the velocity width of the HI emission line.  For the H$\alpha$ velocity widths, only the SFE galaxies are included because even though the K+A definition of \citet{hel07} allows for a low level of H$\alpha$ emission, any detected H$\alpha$ emission is most likely not the result of ongoing star formation.  All of the SFE galaxies with detected HI have stellar velocities significantly greater than $W_{50}$; the median ratio of $W_{50}$(stars) to $W_{50}$ for these galaxies is about 4.4 (see Fig.\ \ref{vwidths}).  All but a few (3-4) K+A galaxies are consistent with the same value for this ratio.  In contrast, the median ratio of $W_{50}$(H$\alpha$) to $W_{50}$ is about 1.5 for the SFE galaxies with three of them having values of $W_{50}$(H$\alpha$) and $W_{50}$ that are essentially the same.  These results imply that it is much more likely that the HI gas is located within the same regions as the star formation rather than throughout the galaxies.  The fact that the HI emission line velocity width tends to be moderately larger than that of the H$\alpha$ emission line may indicate that the neutral hydrogen extends to somewhat larger radii (i.e., where the circular velocity is likely higher) than the emission line gas, or that it extends beyond the area covered by the 3 arcsec aperture used by the SDSS spectrograph.\par
\begin{figure*}
\includegraphics[scale=0.9]{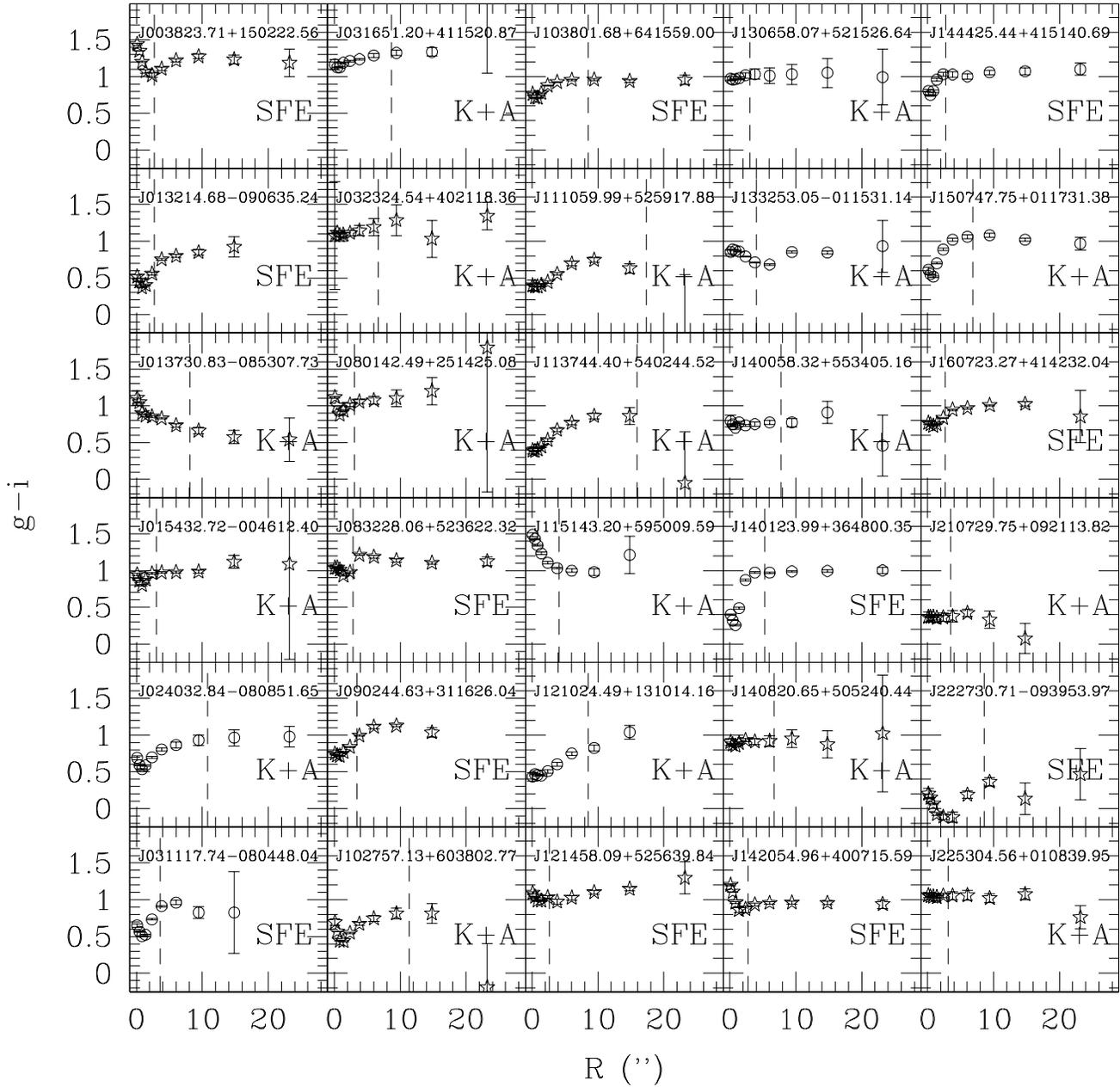}
\caption{The observed $g-i$ profiles taken from the SDSS photometric pipeline \citep{sto02} for all 30 galaxies observed with the GBT.  Galaxies with HI detections are represented by stars; those without detections are represented by open points.  In each panel, the angular size corresponding to 1 kpc for H$_{\circ}$=70 km s$^{-1}$ Mpc$^{-1}$ is marked with a vertical dashed line.}
\label{profs}
\end{figure*}

But, where are the star forming regions located within these galaxies?  To partially answer this question, the $g-i$ surface brightness profiles measured by the SDSS photometric pipeline \citep[see ][]{sto02} within concentric circular apertures for all 30 galaxies are plotted in Fig.\ \ref{profs}.  The majority of the galaxies either have negative $g-i$ gradients within the inner parts of their profiles indicative of increasingly younger stellar populations, or have a "dip" in their profiles most likely due to both a decrease in mean stellar age and an increase in internal dust extinction towards the galaxy centres.  In fact, the $z$-band dust extinction estimates made by \citet{kau03} using model fits to the stellar continua of the SDSS spectra of the galaxies presented here indicate similar levels of dust extinction for the SFE and K+A galaxies.  The $z$-band extinction for both types of galaxies ranges from 0 to $\sim1$ mag with mean values for both classes of about 0.35 mag, corresponding to a colour excess of $E(g-i)\sim0.4$ \citep{sch98}.  Both the negative gradients and the dip features in the surface brightness profiles indicate that for the majority of the galaxies, the star formation is occurring preferentially in the galaxies' centres.  This is similar to what has been found previously for both K+A \citep{yam05} and SFE \citep{helw07} galaxies.  For most of these galaxies, the changes in the $g-i$ profile shapes indicative of star formation occur within the inner kiloparsec, as indicated by the vertical dashed lines in the profiles displayed in Fig.\ \ref{profs}.  When taken into account with the kinematic data plotted in Fig.\ \ref{vwidths}, one would also expect to find the majority of the neutral atomic hydrogen within $\sim$1 kpc from the centres of these galaxies.  Interferometric data obtain with an instrument such as the NRAO Very Large Array is required to produce synthesis images of 21-cm emission with high enough spatial resolution to adequately and properly address this issue.

\section*{Acknowledgements}
The author would like to thank the NRAO TAC and the GBT scheduler for the generous allocation of observing time and for the GBT staff for expert assistance during the observing run.  This research was performed while the author held a National Research Council Research Associateship Award at the Naval Research Laboratory.  Basic research in astronomy at the Naval Research Laboratory is funded by the Office of Naval Research. The National Radio Astronomy Observatory is a facility of the National Science Foundation operated under cooperative agreement by Associated Universities, Inc.  NED is operated by the Jet Propulsion Laboratory, California Institute of Technology, under contract with the National Aeronautics and Space Administration.

\bsp

\label{lastpage}


\begin{thebibliography}{}
\bibitem[\protect\citeauthoryear{Abadi et al.}{1999}]{aba99} Abadi, M. G., Moore, B., \& Bower, R. G. 1999, MNRAS, 308, 947
\bibitem[\protect\citeauthoryear{Aceves \& Vel\'{a}zquez}{2005}]{ace05} Aceves, H. \& Vel\'{a}zquez, H.  2005, MNRAS, 360, L50
\bibitem[\protect\citeauthoryear{Barnes et al.\ }{2001}]{bar01} Barnes, D. G., et al. 2001, MNRAS, 322, 486
\bibitem[\protect\citeauthoryear{Buyle et al.\ }{2006}]{buy06} Buyle, P., Michielsen, D., De Rijcke, S., Pisano, D. J., Dejonghe, H., \& Freeman, K. 2006, ApJ, 649, 163
\bibitem[\protect\citeauthoryear{Chang et al.\ }{2001}]{cha01} Chang, T., van Gorkom, J. H., Zabludoff, A. I., Zaritsky, D., \& Mihos, J. C.  2001, AJ, 121, 1965
\bibitem[\protect\citeauthoryear{Couch \& Sharples}{1987}]{cou87} Couch, W. J. \& Sharples, R. M.  1987, MNRAS, 229, 423
\bibitem[\protect\citeauthoryear{Dressler \& Gunn}{1983}]{dre83} Dressler, A., \& Gunn, J. E.  1983, ApJ, 270, 7
\bibitem[\protect\citeauthoryear{Dressler et al.\ }{1997}]{dre97} Dressler, A., et al. 1997, ApJ, 490, 577
\bibitem[\protect\citeauthoryear{Fasano et al.\ }{2000}]{fas00} Fasano, G., Poggianti, B. M., Couch, W. J., Bettoni, D., Kjaergaard, P., \& Moles, M. 2000, ApJ, 542, 673
\bibitem[\protect\citeauthoryear{Fukugita et al.\ }{2004}]{fuk04} Fukugita, M., Nakamura, O., Turner, E., Helmboldt, J., \& Nichol, R.  2004, ApJL, 601, 127
\bibitem[\protect\citeauthoryear{Goto et al.\ }{2003}]{got03} Goto, T., et al.  2003, PASJ, 55, 771
\bibitem[\protect\citeauthoryear{Goto }{2004}]{got04} Goto, T.  2004, A\&A, 427, 125
\bibitem[\protect\citeauthoryear{Goto }{2005}]{got05} Goto, T.  2005, MNRAS, 357, 937
\bibitem[\protect\citeauthoryear{Gunn \& Gott }{1972}]{gun72} Gunn, J. E., \& Gott, J. R., III. 1972, ApJ, 176, 1
\bibitem[\protect\citeauthoryear{Helmboldt et al.\ }{2004}]{hel04} Helmboldt, J. F., Walterbos, R. A. M., Bothun, G. D., O'Neil, K., \& de Blok, W. J. G. 2004, ApJ, 613, 914
\bibitem[\protect\citeauthoryear{Helmboldt et al.\ }{2007}]{hel07} Helmboldt, J. F., Walterbos, R. A. M., \& Goto, T.  2007, MNRAS, submitted
\bibitem[\protect\citeauthoryear{Helmboldt \& Walterbos }{2007}]{helw07} Helmboldt, J. F. \& Walterbos, R. A. M.  2006, in preparation
\bibitem[\protect\citeauthoryear{Hogg et al.\ }{2006}]{hog06} Hogg, D. W, Masjedi, M., Berlind, A. A., Blanton, M. R., Quintero, A. D., \& Brinkmann, J.  2006, ApJ, 650, 763
\bibitem[\protect\citeauthoryear{Kauffmann et al.\ }{2003}]{kau03} Kauffmann, G., et al.\  2003, MNRAS, 341, 54
\bibitem[\protect\citeauthoryear{Kennicutt }{1998}]{ken98} Kennicutt, R. C. 1998, ARA\&A, 36, 189
\bibitem[\protect\citeauthoryear{Lake \& Schommer }{1984}]{lak84} Lake, G. \& Schommer, R. A.  1984, ApJ, 280, 107
\bibitem[\protect\citeauthoryear{Larson et al.\ }{1980}]{lar80} Larson, R. B., Tinsley, B. M., \& Caldwell, C. N. 1980, ApJ, 237, 692
\bibitem[\protect\citeauthoryear{Leech et al.\ }{1994}]{lee94} Leech, K. J., Rowan-Robinson, M., Lawrence, A., \& Hughes, J. D.  1994, MNRAS, 267, 253
\bibitem[\protect\citeauthoryear{Meza et al.\ }{2003}]{mez03} Meza, A., Navarro, J. F., Steinmetz, M., \& Eke, V. R.  2003, ApJ, 590, 619
\bibitem[\protect\citeauthoryear{Nulsen }{1982}]{nul82} Nulsen, P. E. J. 1982, MNRAS, 198, 1007
\bibitem[\protect\citeauthoryear{Pogge \& Eskridge }{1993}]{pog93} Pogge, R. W. \& Eskridge, P. B.  1993, AJ, 106, 1405
\bibitem[\protect\citeauthoryear{Quintero et al.\ }{2004}]{qui04} Quintero, A. D. et al.  2004, ApJ, 602, 190
\bibitem[\protect\citeauthoryear{Sanders \& Mirabel }{1996}]{san96} Sanders, D. B. \& Mirabel, I. F.  1996, ARA\&A, 34, 749
\bibitem[\protect\citeauthoryear{Schlegel et al.\ }{1998}]{sch98} Schlegel, D. J., Finkbeiner, D. P., \& Davis, M.  1998, ApJ, 500, 525
\bibitem[\protect\citeauthoryear{Smith et al.\ }{2002}]{smi02} Smith, J. A., et al.  2002, AJ, 123, 2121
\bibitem[\protect\citeauthoryear{Stoughton et al.\ }{2002}]{sto02} Stoughton, C., et al.  2002, AJ, 123, 485
\bibitem[\protect\citeauthoryear{van Dokkum et al.\ }{1998}]{van98} van Dokkum, P. G., Franx, M., Kelson, D. D., Illingworth, G. D., Fisher, D., \& Fabricant, D. 1998, ApJ, 500, 714
\bibitem[\protect\citeauthoryear{Tran et al.\ }{2004}]{tra04} Tran, K. H., Franx, M., Illingworth, G. D., van Dokkum, P., Kelson, D. D., \& Magee, D.  2004, ApJ, 609, 683
\bibitem[\protect\citeauthoryear{Tremonti et al.\ }{2004}]{tre04} Tremonti, C. A., et al.\  2004, ApJ, 613, 898
\bibitem[\protect\citeauthoryear{Vogt et al.\ }{2004}]{vog04} Vogt, N. P., Haynes, M. P., Giovanelli, R., \& Herter, T.  2004, AJ, 127, 3300
\bibitem[\protect\citeauthoryear{Worthey \& Ottaviani }{1997}]{wor97} Worthey, G. \& Ottaviani, D. L.  1997, ApJS, 111, 377
\bibitem[\protect\citeauthoryear{Yamauchi \& Goto }{2005}]{yam05} Yamauchi, Y. \& Goto, T.  2005, MNRAS, 359, 1557
\bibitem[\protect\citeauthoryear{York et al.\ }{2000}]{yor00} York, D. et al.  2000, AJ, 120, 1579
\bibitem[\protect\citeauthoryear{Zabludoff et al.\ }{1996}]{zab96} Zabludoff, A. I., Zaritsky, D., Lin, H., Tucker, D., Hashimoto, Y., Shectman, S. A., Oemler, A., \& Kirshner, R. P.  1996, ApJ, 466, 104
\end{thebibliography}
\end{document}